# Preprocessing: A Prerequisite for Discovering Patterns in Web Usage Mining Process

Ramya C, Dr. Shreedhara K S and Kavitha G

M.Tech (Final Year), Professor & Chairman and Lecturer, Dept. of Studies in CS&E
U.B.D.T College of Engineering, Davangere
Davangere University, Karnataka, INDIA
cramyac@gmail.com and ks_shreedhara@yahoo.com

*Abstract*—Web log data is usually diverse and voluminous. This data must be assembled into a consistent, integrated and comprehensive view, in order to be used for pattern discovery. Without properly cleaning, transforming and structuring the data prior to the analysis, one cannot expect to find meaningful patterns. As in most data mining applications, data preprocessing involves removing and filtering redundant and irrelevant data, removing noise, transforming and resolving any inconsistencies. In this paper, a complete preprocessing methodology having merging, data cleaning, user/session identification and data formatting and summarization activities to improve the quality of data by reducing the quantity of data has been proposed. To validate the efficiency of the proposed preprocessing methodology, several experiments are conducted and the results show that the proposed methodology reduces the size of Web access log files down to 73-82% of the initial size and offers richer logs that are structured for further stages of Web Usage Mining (WUM). So preprocessing of raw data in this WUM process is the central theme of this paper.

## I. INTRODUCTION

The number of Web resources are available on the Internet as well as the number of Web users is continuously growing. As a result, the quantity of the usage data available for a WUM study is also increasing. The most common Web log format is CLF/ECLF shown in Fig. 3 & Fig. 4. In this paper, we describe a general methodology for preprocessing the raw web logs into a structured form. A log file is a plain text file, where requests are ordered chronologically by the time at which the user requested the resource. Usually, a data mining tool needs records as input, stored as rows in a database table or as transactions (i.e. sequences of items). Fig. 1 shows the web usage mining in which preprocessing is the first stage wherein the raw web logs are preprocessed into a structured form.

The aims of the preprocessing step in a WUM process are roughly to convert the raw log file into a set of transactions (one transaction being the list of pages visited by one user) and to discharge the non-interesting or noisy requests (e.g. implicit requests or requests made by Web robots). The main objectives of preprocessing are to reduce the quantity of data being analyzed while, at the same time, to enhance its quality. The results show that the proposed methodology reduces the size of Web access log files down to 73-82% of the initial size and offers richer logs that are structured for further stages of Web Usage Mining.

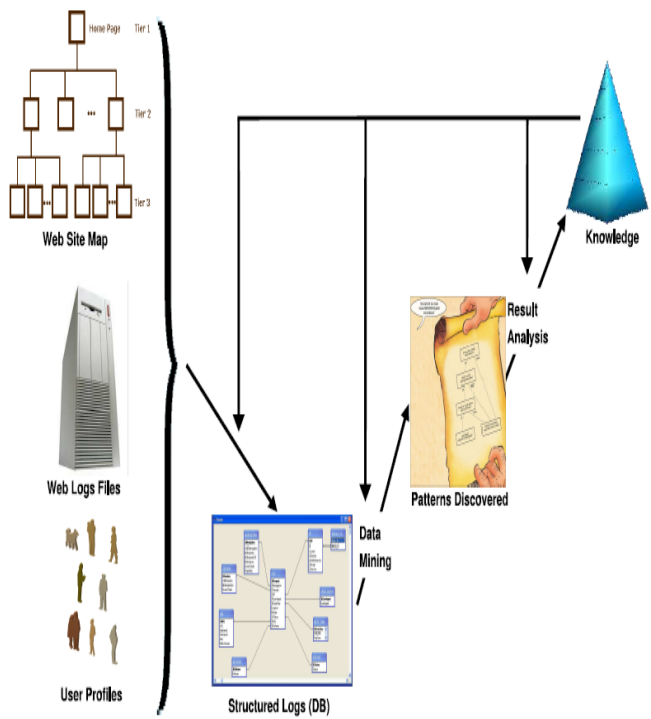

Figure 1. General Web Usage Mining Process

```
72.30.252.91 - - [18/Jun/2006:12:28:33 +0000] "GET
/robots.txt HTTP/1.0" 200 52 "-" "Mozilla/5.0
(compatible; Yahoo! Slurp;
http://help.yahoo.com/help/us/ysearch/slurp)"
83.77.134.184 - - [18/Jun/2006:12:29:40 +0000] "GET
/vanuatu/export/system/modules/VTO/resources/style
sheet/vto.css HTTP/1.1" 200 7797 "-" "Mozilla/4.0
(compatible; MSIE 6.0; Windows NT 5.1; SV1; .NET
CLR 1.1.4322)"
83.77.134.184 - - [18/Jun/2006:12:29:41 +0000] "GET
/vanuatu/export/sites/VTO/fr/kids/volcanoes/ambrym
_eruption.html HTTP/1.1" 200 26812 "-"
"Mozilla/4.0 (compatible; MSIE 6.0; Windows NT
5.1; SV1; .NET CLR 1.1.4322)"
83.77.134.184 - - [18/Jun/2006:12:29:41 +0000] "GET
/vanuatu/export/system/modules/VTO/resources/ima
ges/nto_kids_logo.jpg HTTP/1.1" 200 10420 "-"
"Mozilla/4.0 (compatible; MSIE 6.0; Windows NT
5.1; SV1; .NET CLR 1.1.4322)"
83.77.134.184 - - [18/Jun/2006:12:29:41 +0000] "GET
/vanuatu/export/system/modules/VTO/resources/ima
ges/vanuatu.gif HTTP/1.1" 200 40892 "-"
"Mozilla/4.0 (compatible; MSIE 6.0; Windows NT
5.1; SV1; .NET CLR 1.1.4322)"
```

Figure 2. Access log of the web server

```
<ip_addr><base_url><date>
<method><file><protocol>
<code><bytes>
```

Figure 3. Common Log Format (CLF)

```
<ip_addr><base_url><date><method>
<file><protocol><code><bytes>
<referrer>
```

Figure 4. Extended Common Log Format (ECLF)

## II. PREPROCESSING

Web log data is usually diverse and voluminous. This data must be assembled into a consistent, integrated and comprehensive view, in order to be used for pattern discovery. As in most data mining, data preprocessing involves removing noise, transforming and resolving any inconsistencies. For example, requests for graphical page content and requests for any other file which might be included into a Web page or even navigation sessions performed by robots and Web spiders are removed. Without properly cleaning, transforming and structuring the data prior to the analysis one cannot expect to find the meaningful patterns. The information provided by the data sources can be used to construct a data model consisting of several data abstractions. Web servers are surely the richest and most common source of data. They can collect large amount of data from the Web site's. The data is stored in the Web access log files. A typical example of web log file is shown in Fig. 2. Each access to a Web page is recorded in the access log of the Web server that hosts it. The entries of a Web log file consist of fields that follow a predefined format such as Common Log Format (CLF), Extended Common Log Format (ECLF). A CLF file is created by the Web server to keep track of the requests that occur on a Web site. ECLF is supported by Apache and Netscape (W3C). The CLF and ECLF are shown in Fig. 3 & Fig. 4.

The Stages of Preprocessing are shown in Fig. 5. It comprises of the following steps – Merging of Log files from Different Web Servers, Data cleaning, Identification of Users, Sessions, and Visits, Data formatting and Summarization.

*A. Merging*

At the beginning of the data preprocessing, the requests from all log files in *Log*, put together into a joint log file '*£*' with the Web server name to distinguish between requests made to different Web servers and taking into account the synchronization of Web server clocks, including time zone differences. The merging problem is formulated as "Given the set of log files $Log = \{L_1, L_2, L_3...L_n\}$, merge these log files into a single log file *£ (joint log file)*". Let $L_i$ be the $i^{th}$ log file. Let $L_{i.c}$ is a cursor on $L_i$'s requests and $L_{i.l}$ is the current log entry from $L_i$ indicated by $L_{i.c}$. Let $L_{i.l.time}$ be the time t of the current log entry of $L_i$. Let $S=(w_1,w_2,...w_n)$ is an array with Web server names, where S[i] is the Web server's name for the log $L_{i.l}$.

*Steps:*
*1. Initialize the joint log file £ cursor*
*2. Scan the log entries from each log file $L_i$ in Log and append to £*
*3. Sort the £ entries in ascending order based on access time*
*4. Return £*

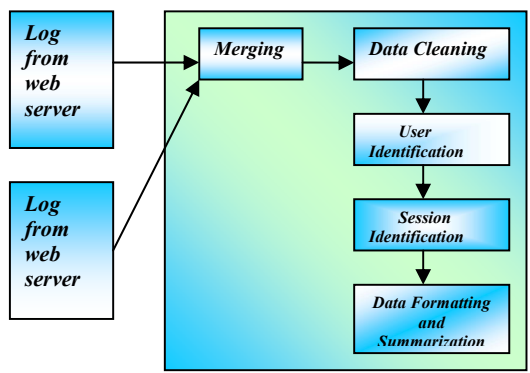

Figure 5. Stages of Preprocessing

*B. Data Cleaning*

The second step of data preprocessing consists of removing useless requests from the log files. Since all the log entries are not valid, we need to eliminate the irrelevant entries. Usually, this process removes requests concerning non-analyzed resources such as images, multimedia files, and page style files. For example, requests for graphical page content (*.jpg & *.gif images) and requests for any other file which might be included into a web page or even navigation sessions performed by robots and web spiders. By filtering out useless data, we can reduce the log file size to use less storage space and to facilitate upcoming tasks.

*C. User Identification*

In most cases, the log file provides only the computer address (name or IP) and the user agent (for the ECLF log files). For Web sites requiring user registration, the log file also contains the user login (as the third record in a log entry) that can be used for the user identification. When the user login is not available, each IP is considered as a user, although it is a fact that an IP address can be used by several users. For Knowledge Discovery from Web Usage Data (KDWUD), to get knowledge about each user's identity is not necessary. However, a mechanism to distinguish different users is still required for analyzing user access behavior.

*D. Session Identification*

A user session is a directed list of page accesses performed by an individual user during a visit in a Web site. i.e., the group of activities performed by a user from the moment he/she entered the site to the moment he/she left it.

A user may have a single (or multiple) session(s) during a period of time. Thus the session identification problem is formulated as "Given the Web log file *Log,* capture the Web users' navigation trends, typically expressed in the form of Web users' sessions". Fig. 6 summarizes the session identification process with a pseudo code. *'Session_Gen'* function calls the *'Distance'* function. Given a list of histories and a page file *f*, the *'Distance'* function finds the history that most recently accessed *f*.

```
Session_Gen ()
{ Let H_i={f_1,f_2,..., f_n} denote the time-ordered session
  history
  Let l_j, f_j, r_j, and t_j denote log entry, request, referrer, and
  time  (at which the request was received) respectively.
  Let τ denote the session time-out (Usually 30 minutes)
  Sort the Log data by IP address, agent, and time.
  for each unique IP/agent combinations do  {
      for each l_j do  {
          if ( ( (t_j-t_{j-1}) > τ ) or ( r_j ∉ {H_0,H_1,..., H_m}) )
              then increment i, add l_j to H_i
          else    assign = Distance (H, r_j), add l_j to H_{assign}
      }
  }}
```

Figure 6. Pseudo code for session identification

*E. Data Formatting & Summarization*

This is the last step of data preprocessing. Here, the structured file containing sessions and visits are transformed to a relational database model. Then, the data generalization method is applied at the request level and aggregated for visits and user sessions to completely fill in the database. The data summarization concerns with the computation of aggregated variables at different abstraction levels (e.g. request, visit, and user session). These aggregated variables are later used in the data mining step. They represent statistical values that characterize the objects analyzed. For instance, if the object analyzed is a **user session**, in the aggregated data computation process, the following variables are calculated.

- The number of visits for that session
- The length of the session in seconds (the difference between the last and the first date of the visit) or in pages viewed (the total number of page views)
- The number of visits for the period considered, which can be a day, a week, or a month

Similarly, other aggregated variables that can be computed are:

- The percentage of the requests made to each Web server.
- Number of unique visitors/hosts per hr/day/week/month
- Number of unique user agents per hr/day/week/month

Depending on the objective of the analysis, the analyst can decide to compute additional and more complex variables.

## III. EXPERIMENTAL RESULTS

We have conducted several experiments on log files collected from NASA Web site during July 1995. Through these experiments, we show that our preprocessing methodology reduces significantly the size of the initial log files by eliminating unnecessary requests and increases their quality through better structuring. This is shown in Table 1. It is observed from the Table 1 that, the size of the log file is reduced to 73-82% of the initial size. The Fig. 7 shows the GUI of our toolbox with preprocessor tab. The user sessions are shown in the Table 2.

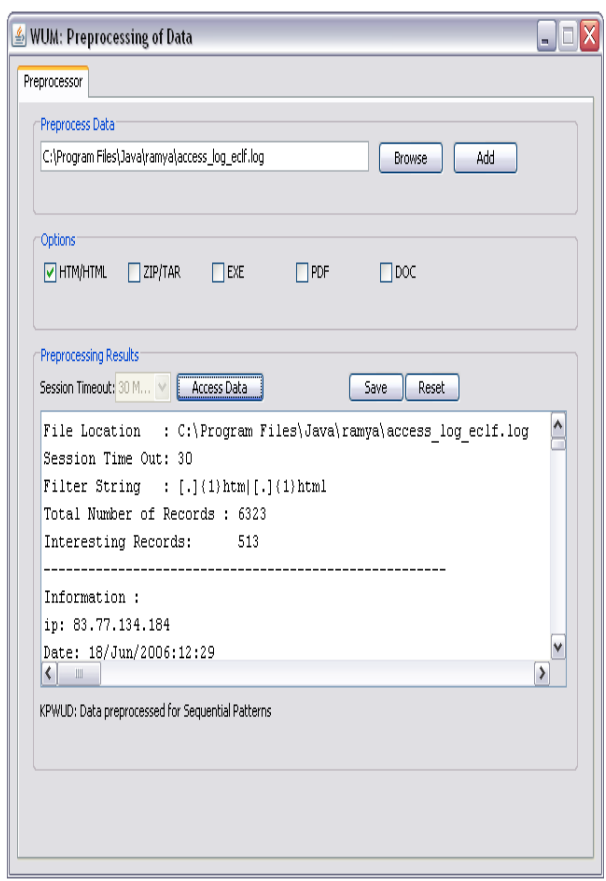

Figure 7. WUM toolbox: Results after preprocessing (NASA Log file, Aug 1995)

TABLE I. RESULTS AFTER PREPROCESSING

| Website | Duration | Original Size | Size after Preprocessing | % Reduction in Size | No. of Sessions | No. of Users |
|---|---|---|---|---|---|---|
| NASA | 1-10th Aug 95 | 75361 bytes (7.6MB) | 20362 bytes | 72.98% | 6821 | 5421 |
| NASA | 20-24th July 95 | 205532 bytes (20.6MB) | 57092 bytes | 72.22% | 16810 | 12525 |
| Academic Site | 12-28th May 2001 | 28972 bytes (2.9MB) | 5043 bytes | 82.5% | 1645 | 936 |

## IV. COMPARISON OF PREPROCESSING METHODOLOGIES

After illustrating the use of our preprocessing methodology through different processes, we present in this section the main related works in this domain. In the recent years, there has been much research on Web usage mining [3,4,5,6,7,8,9,10,11,12,13,14,15,16]. However, as described below, data preprocessing in WUM has received far less attention than it deserves. Methods for user identification, sessionizing, page view identification, path completion, and episode identification are presented in [3]. However, some of the heuristics proposed are not appropriate for larger and more complex Web sites. We compared our preprocessing methodology with the preprocessing described in other general WUM research works [3, 9, 10, 11, 12, 15 and 16].The results of this comparison are provided in Table 3. The comparison Table 3 focuses only on the data preprocessing step and it shows how different preprocessing features were implemented in the main related works.

## V. CONCLUSIONS

WUM is intended for Web site authors and administrators who want to improve the organization of their Web documents and adapt it better to the needs of the information consumers. The proposed preprocessing methodology is more complete because it offers the possibility of analyzing jointly multiple Web server logs. It employs effective heuristics for detecting and eliminating Web robot requests. It proposes a relational database model for storing the structured information about the Web site, its usage and its users with many aggregations.

## VI. FUTURE WORK

There are a number of unsolved technical problems and open issues at the stage of data collection & preprocessing. New techniques and possibly new models for acquiring data are needed. A Poll by KDnuggets (15/3/2000 – 30/3/2000) revealed that, about 70% of the users consider Web mining as a compromise of their privacy. Thus it is imperative that new Web usage mining tools are transparent to the user by providing access to the data collected and clarifying the use of these data as well as the potential benefits for the user. Preprocessing methodology can be further extended by using site maps and semantic topics of Web pages for page view and episode identification.

TABLE II. USER SESSIONS

| Session Id | IP Address | Date &Time | URL Accessed |
|---|---|---|---|
| 9 | 128.102.204.243 | 1995-07-22 01:16:58 | /shuttle/missions/sts-73/mission-sts-73.html |
| 9 | 128.102.204.243 | 1995-07-22 01:17:25 | /shuttle/missions/sts-74/mission-sts-74.html |
| 9 | 128.102.204.243 | 1995-07-22 01:17:38 | /shuttle/missions/sts-72/mission-sts-72.html |
| 9 | 128.102.204.243 | 1995-07-22 01:17:45 | /shuttle/missions/sts-75/mission-sts-75.html |
| 9 | 128.102.204.243 | 1995-07-22 01:17:52 | /shuttle/missions/sts-76/mission-sts-76.html |
| 9 | 128.102.204.243 | 1995-07-22 01:17:58 | /shuttle/missions/sts-77/mission-sts-77.html |
| 9 | 128.102.204.243 | 1995-07-22 01:18:05 | /shuttle/missions/sts-78/mission-sts-78.html |
| 10 | 128.102.210.40 | 1995-07-20 23:27:49 | /shuttle/countdown/countdown.html |
| 10 | 128.102.210.40 | 1995-07-20 23:28:11 | /shuttle/technology/sts-newsref/stsref-toc.html |
| 10 | 128.102.210.40 | 1995-07-20 23:28:57 | /shuttle/technology/sts-newsref/sts_mes.html |
| 10 | 128.102.210.40 | 1995-07-20 23:29:11 | /shuttle/countdown/liftoff.html |
| 10 | 128.102.210.40 | 1995-07-20 23:30:18 | /shuttle/missions/sts-69/mission-sts-69.html |
| 11 | 128.102.210.40 | 1995-07-21 01:58:47 | /shuttle/countdown/countdown.html |
| 11 | 128.102.210.40 | 1995-07-21 01:59:12 | /shuttle/countdown/liftoff.html |

TABLE III. COMPARISON OF PREPROCESSING METHODOLOGIES

| Related work | Data Source | | | Data Cleaning | | | | Data Formatting & Structuring | | | |
|---|---|---|---|---|---|---|---|---|---|---|---|
| | Site Map | Site Semantic | Multiple Servers | Merging | Anonym-ization | Removing Images | Removing Web Robots | User ID | Session ID | Visit ID | Generalization & Aggregation |
| Berendt [15] | | | | | | √ | | | IP | $H_{Visit}$ | |
| Chen [12] | √ | | | | | | | | | | |
| Cooley [3] | √ | | | | √ | √ | √ | Login | IP, Agent | $H_{Visit}$ $H_{Ref}$ | |
| Fu [11] | | | | | | √ | | | IP | $H_{Visit}$ | √ |
| Krishnapuram [9] | | | | | | √ | | | IP | | |
| Shahabi [10] | √ | | | | √ | | | | IP, Session ID | $H_{Visit}$ $H_{Page}$ | |
| Becker [16] | | | | | | √ | | Login | Login | $H_{Ref}$ | |
| Our Method | | | √ | √ | √ | √ | √ | IP,OS Agent | IP, Agent | $H_{Visit}$ $H_{Page}$ | √ |